\documentclass[prb,twocolumn,showpacs,preprintnumbers,amsmath,amssymb]{revtex4}

\usepackage{graphicx}
\usepackage{dcolumn}
\usepackage{bm}

\begin{document}

\title{Plasmon tunability in metallodielectric metamaterials}

\author{Sampsa Riikonen,$^1$ Isabel Romero,$^1$ and F. J. Garc\'{\i}a de Abajo$^{1,2}$\cite{correspondingauthor}}
\affiliation{$^1$Donostia International Physics Center, Paseo Manuel
de Lardizabal 4, 20018 San Sebastian, Spain \\ $^2$Unidad de
F\'{\i}sica de Materiales CSIC-UPV/EHU Aptdo. 1072, 20080 San
Sebastian, Spain}
\date{\today}

\begin{abstract}
The dielectric properties of metamaterials consisting of
periodically arranged metallic nanoparticles of spherical shape
are calculated by rigorously solving Maxwell's equations.
Effective dielectric functions are obtained by comparing the
reflectivity of planar surfaces limiting these materials with
Fresnel's formulas for equivalent homogeneous media, showing
mixing and splitting of individual-particle modes due to
inter-particle interaction. Detailed results for simple cubic and
fcc crystals of aluminum spheres in vacuum, silver spheres in
vacuum, and silver spheres in a silicon matrix are presented. The
filling fraction of the metal $f$ is shown to determine the
position of the plasmon modes of these metamaterials. Significant
deviations are observed with respect to Maxwell-Garnett effective
medium theory for large $f$, and multiple plasmons are predicted
to exist in contrast to Maxwell-Garnett theory.
\end{abstract}
\maketitle

\section{Introduction}

Artificial materials with tailored optical response functions
(metamaterials) constitute a rich field of research due to their
applicability to facilitate the design of highly-demanding optical
devices for computing and communications technologies. That is the
case of left-handed materials, which posses negative index of
refraction \cite{SSS01} and permit designing lenses of resolution
below the diffraction limit. \cite{P00} Furthermore, newly
available nanostructured materials \cite{ULM01,PKA01,sam1,DKE04}
pose interesting possibilities that deserve detailed exploration
to search for extraordinary optical properties.

Composite materials with components of dimensions much smaller
than the light wavelength can be assimilated in general to
equivalent homogeneous media with macroscopic optical properties
such as a dielectric function and a magnetic permeability. The
problem of finding the effective optical response of arrays of
homogeneous spheres was already discussed by Maxwell. \cite{M1873}
In fact, effective medium theories that yield simple analytical
expressions have been available for nearly a century, like in the
case of Maxwell-Garnett's theory (MGT), \cite{M1904} which is
particularly suited to describe spherical inclusions and rather
accurate at small concentrations of the latter $f$, or Bruggeman's
theory, \cite{B1935} less accurate but intended to work for
arbitrary values of $f$. Extensions and corrections to these
theories have been reported over the last decades for periodic
\cite{B1979,S1980,LWA1980,M1981,TCS1990,BD92,DCH93,MP95,LDL00,LCS00,MZT00,paper075,IPK04}
and disordered \cite{LWA1980,CKK1990,SKC1990,HF92,BF95,NB98,LDL00}
composites, all of them sharing in common the assumption of local
response.

For metallic components plasmons are observed as collective
oscillations of the valence electrons. In bulk metals, they are
signalled by the vanishing of bulk dielectric function,
$\epsilon=0$. For an ideal metallic behavior as that described by
Drude's formula
\begin{eqnarray}
  \epsilon(\omega)=1-\frac{\omega_p^2}{\omega(\omega+{\rm i}
  \eta)},
  \label{Drude}
\end{eqnarray}
with $\eta\rightarrow 0^+$, this occurs at the bulk plasmon
frequency $\omega=\omega_p$. The plasmon frequency is affected by
the shape of the boundary of the metal, and for instance surface
plasmons occur with frequency $\omega_s=\omega_p/\sqrt{2}$ at a
planar surface, whereas modes of frequencies $\omega_l=\omega_p
\sqrt{l/(2l+1)}$ are observed in spherical particles in the
long-wavelength limit, corresponding to multipole oscillations of
orders $l=1,2,\dots$, as described in the early work of Mie.
\cite{M1908} Multipolar plasmons can be also found in
non-spherical metallic particles like nanorods, \cite{KSR00}
nanorings, \cite{paper072} and nanoshells. \cite{PRH03} Collective
modes in these particles and in particle pairs have been explained
intuitively by invoking {\it plasmon chemistry} concepts following
the general procedure outlined in Ref. \onlinecite{NL04}. For
layers of particles on a substrate, new plasmons are observed as a
result of their mutual interaction. \cite{TRC99} In 3D periodic
arrangements of spheres like the ones discussed below, Inglesfield
{\it et al.} \cite{IPK04} have reported a comprehensive study of
plasmon bands and have found that wide bands do exist and that
significant deviations from MGT occur in near-touching metal
spheres. This connects to previous results on plasmon-like
behavior in wired structures at THz frequencies. \cite{PHS96}

In this work we concentrate on metamaterials formed by periodic
arrays of metallic spherical nanoparticles confined within a host
dielectric medium. We calculate their effective dielectric
function from a rigorous solution of Maxwell's equations,
therefore including all multipole corrections in an exact fashion,
as a function of metallic filling fraction $f$. Tunable plasmons
like the ones under study here can find application to light
sensors, absorbers and emitters of light, and functional lasers.
\cite{SEB02} The relevance of this study is illustrated by the
several methods that have been developed for synthesizing
nanocomposites, making use of nanoparticles self-assembly
\cite{ULM01,ZMU99,GPG01,FYB04} to prepare ordered layers of
metallic particles in dielectric hosts, including fcc arrangements
of Au particles in Si \cite{FYB04} and other complex particle
shapes. \cite{PKA01,sam1}

This paper shows how the plasma frequencies can be engineered in
metallodielectric composites consisting of spherical metallic
nanoparticles embedded in a dielectric host. We consider in
particular simple cubic (sc) and fcc arrays of aluminium
nanoparticles in vacuum, as well as fcc arrays of silver
nanoparticles surrounded by either vacuum or silicon. Modes other
than dipolar are not resolved by external light in isolated
nanoparticles, which leads to the prediction of a single plasmon
branch as a function of metallic filling fraction when MGT is
used. However, our rigorous solution of Maxwell's equations
reveals several plasmon branches as a result of mixing and
splitting of individual-particle modes of multipolar nature.

\section{Plasmon chemistry}

\begin{figure}
\includegraphics[keepaspectratio,width=6.0cm]{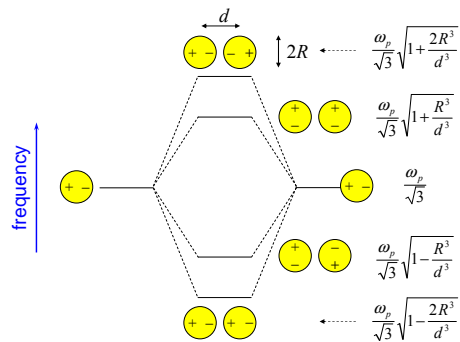}
\caption{\label{Fig1} Plasmon chemistry in interacting
nanoparticles. The non-retarded plasmon energies of two identical
metallic particles are shown along with schematic representations of
the corresponding oscillation modes. The particles have radius $R$,
their center-to-center distance is $d$, and the metal is described
by a Drude dielectric function of bulk-plasma frequency $\omega_p$
[Eq.\ (\ref{Drude})].}
\end{figure}

The basic ingredients of the composites under consideration are
metallic nanoparticles. A isolated metallic nanoparticle can
exhibit dipolar excitations with triple degeneracy (along the
three orthogonal directions of space). When the dielectric
function of the metal is described by Eq.\ (\ref{Drude}), the
frequency of these modes is $\omega_1=\omega_p/\sqrt{3}$.
\cite{sam2} However, the interaction between neighboring
nanoparticles can produce mixing and splitting of these dipolar
modes, in a way similar to the mixing and splitting of p orbitals
in molecular binding, which suggests the term of {\it plasmon
chemistry}. The mode frequencies and their corresponding
polarization patterns for two spheres of radius $R$ interacting at
a certain distance $d$ are represented in Fig.\ \ref{Fig1}, where
the two intermediate frequency states are doubly degenerate. Of
these modes, only the lowest lying one and the third one
($\omega_1 \sqrt{1+R^3/d^3}$) display a net dipole moment, and
therefore, external light with these two frequencies should couple
strongly to this structure, whereas the other modes are either
symmetry forbidden, depending on the light incidence conditions,
or contribute very little to light scattering (silent modes).

\begin{figure}
\includegraphics[keepaspectratio,width=6.0cm]{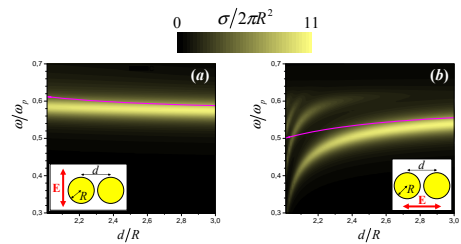}
\caption{\label{Fig2} Total scattering cross section $\sigma$ of a
system formed by two metallic spheres as a function of incident
photon energy $\omega$ and center-to-center distance $d$ between two
aluminum spheres described by the Drude dielectric function of Eq.\
(\ref{Drude}) with $\omega_p=15$ eV and $\eta=0.6$ eV. Two different
polarization directions of the external electric field have been
considered in (a) and (b), as shown in the insets. $\sigma$ is
normalized to the projected area of the two spheres, $2\pi R^2$. The
solid curves represent the modes given by the expressions of Fig.\
\ref{Fig1}.}
\end{figure}

These conclusions are corroborated by the scattering cross section
represented in Fig.\ \ref{Fig2} for two small Al spheres as a
function of light frequency and distance between the sphere
centers $d$. Two different polarization directions have been
considered, as shown in the insets. Results similar to those of
Figs.\ \ref{Fig1} and \ref{Fig2} have been reported in Ref.\
\onlinecite{NL04}.

The cross section is obtained by using the multiple elastic
scattering of multipole expansions method (MESME), designed to
solve Maxwell's equations in the presence of large numbers of
particles in arbitrary positions. \cite{paperMESME} In it, the
electric field is expressed in frequency space $\omega$ in terms
of magnetic and electric scalar functions $\psi_i^M$ and
$\psi_i^E$, respectively, as \cite{L97}
   \begin{eqnarray}
      {\bf E}=   \sum_i [{\bf L}_i\psi_i^M
           - \frac{{\rm i}}{k} \nabla\times{\bf L}_i\psi_i^E],
   \nonumber
   \end{eqnarray}
where $k=\omega/c$, ${\bf L}_i=-{\rm i}({\bf r}-{\bf
r}_i)\times\nabla$ is the orbital angular-momentum operator, the
sum is extended over particle positions ${\bf r}_i$, and $M$ and
$E$ refer to magnetic and electric polarization, respectively. The
electromagnetic field created by some external source is expanded
in terms of spherical waves as
   \begin{eqnarray}
      \psi_i^{\rm ext}({\bf r}) = \sum_{lm} {\rm i}^l
                                   j_l(k|{\bf r}-{\bf r}_i|)
                                   Y_{lm}(\widehat{{\bf r}-{\bf r}_i})
                                   \, \psi_{i,lm}^{\rm ext},
   \nonumber
   \end{eqnarray}
where $j_l$ is a spherical Bessel function, $Y_{lm}$ is a
spherical harmonic, and a polarization index (either magnetic or
electric) is understood. In the absence of multiple scattering,
the induced part of the scalar functions $\psi_i^{\rm
ind}=\psi_i^{\rm ss}$ results from single scattering (ss) of
$\psi_i^{\rm ext}$ by each object:
   \begin{eqnarray}
      \psi_i^{\rm ss}({\bf r}) = \sum_{lm} {\rm i}^l
                    h_l^{(+)}(k|{\bf r}-{\bf r}_i|)
                    Y_{lm}(\widehat{{\bf r}-{\bf r}_i})
                    \, \psi^{\rm ss}_{i,lm},
   \label{eq4}
   \end{eqnarray}
for ${\bf r}$ outside the particle centered at ${\bf r}_i$. Here,
$h_l^{(+)}$ is a spherical Hankel function. The relation between
$\psi_i^{\rm ext}$ and $\psi_i^{\rm ss}$ is provided by the
scattering matrix $t_i$, implicitly defined by $\tilde{\psi}^{\rm
ss}_i = t_i \tilde{\psi}^{\rm ext}_i$, where $\tilde{\psi}^{{\rm
ss}({\rm ext})}_i$ is a vector of components $\psi^{{\rm ss}({\rm
ext})}_{i,lm}$. The matrix $t_i$ describes the full scattering
properties of object ${\bf r}_i$. In a cluster of several
particles, $\psi^{\rm ind}$ takes the same form as Eq.\
(\ref{eq4}), except that it is made up of $\psi^{\rm ss}_i$
(single scattering of the external field at ${\bf r}_i$) plus the
result of the free propagation of $\psi^{\rm ind}_j$ from each
object ${\bf r}_j\neq{\bf r}_i$, followed by scattering at ${\bf
r}_i$ (self-consistent multiple scattering). That is,
\cite{paperMESME}
   \begin{eqnarray}
      \tilde{\psi}^{\rm ind}_i =
             \tilde{\psi}_i^{\rm ss} +
             t_i \sum_{j\neq i}
                          H_{ij}\tilde{\psi}^{\rm ind}_j,
   \label{eq9}
   \end{eqnarray}
where the operator $H_{ij}$ describes the noted propagation. In
this work we consider homogeneous spheres, so that $t_{i}$ becomes
diagonal and given by analytical expressions. \cite{paperMESME}
Furthermore, we have found convergent results, using a finite
number of multipoles with $l\leq 8$.

For a bi-sphere system, Fig.\ \ref{Fig2}(b) demonstrates that only
the lowest-frequency mode is excited for incident-light
polarization along the inter-particle direction, whereas Fig.\
\ref{Fig2}(a) shows that only the third hybridized dipolar mode
contributes for polarization perpendicular to the inter-particle
direction, as expected from the above discussion. The plasmon
modes derived from the simple dipole-mode hybridization model of
Fig.\ \ref{Fig1} follow the regions of large cross section only
for relatively large values of $d$ (solid curves in Fig.\
\ref{Fig2}). Moreover, new resonances show up for parallel
polarization [Fig.\ \ref{Fig2}(b)] when the spheres are close
together, as a result of stronger interaction among multipoles of
the spheres that produce hybridized modes with net dipole moment
along the external field direction.

\begin{figure}
\includegraphics[keepaspectratio,width=6.0cm]{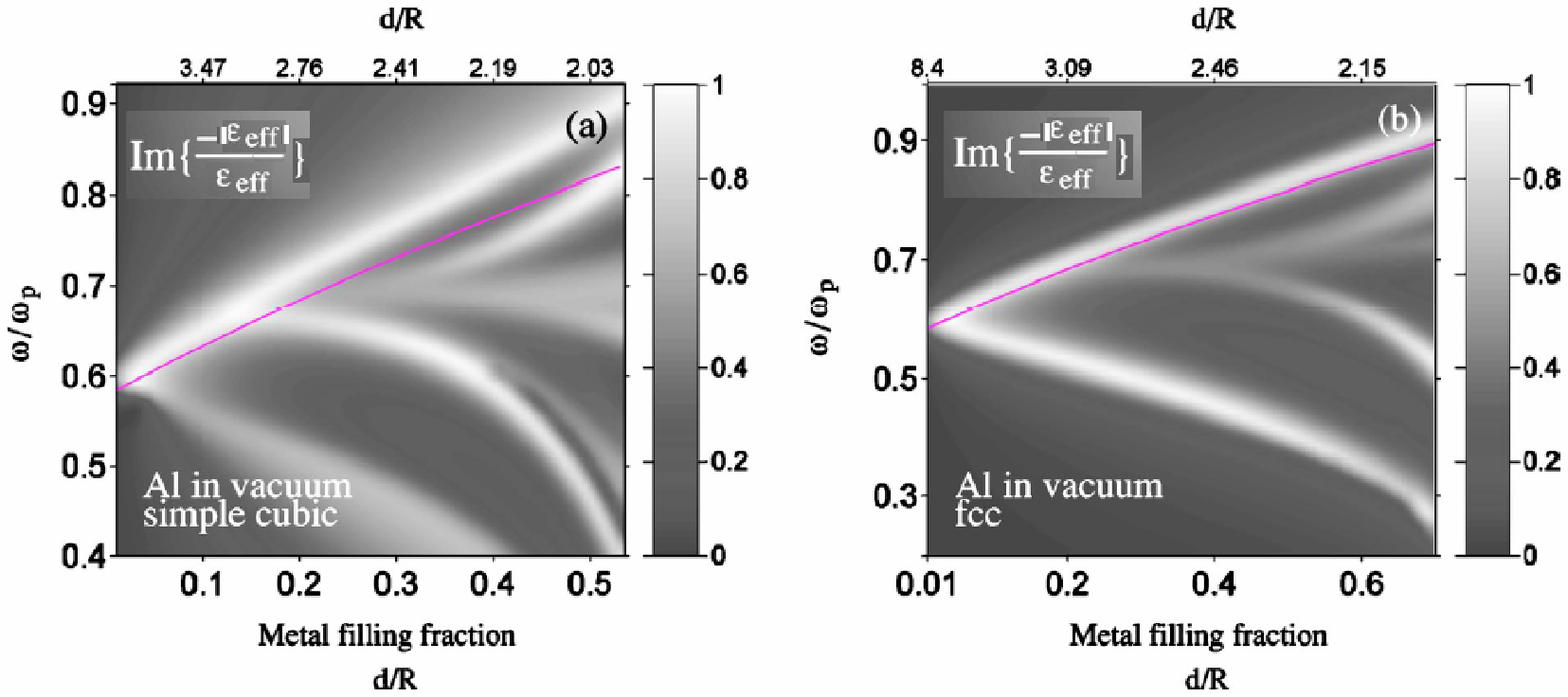}
\caption{\label{Fig3} Contour plot of ${\rm Im}\{-|\epsilon_{\rm
eff}|/\epsilon_{\rm eff}\}$ as a function of incident photon energy
$\omega$ and filling fraction of the metal for aluminium spheres in
simple cubic (a) and fcc (b) configurations, surrounded by vacuum.
Brighter regions correspond to higher values of ${\rm
Im}\{-|\epsilon_{\rm eff}|/\epsilon_{\rm eff}\}$. The solid curves
correspond to Maxwell-Garnett theory as given by Eq.\
(\ref{plasmonlines}).}
\end{figure}

\begin{figure}
\includegraphics[keepaspectratio,width=6.0cm]{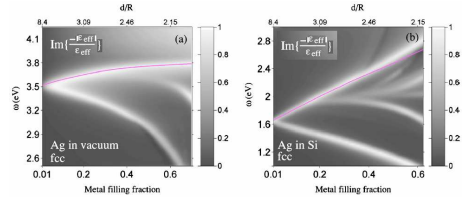}
\caption{\label{Fig4} Contour plot of ${\rm Im}\{-|\epsilon_{\rm
eff}|/\epsilon_{\rm eff}\}$ as a function of incident photon energy
$\omega$ and filling fraction of the metal for silver spheres in an
fcc configuration, surrounded by (a) vacuum or (b) silicon. Brighter
regions correspond to higher values of ${\rm Im}\{-|\epsilon_{\rm
eff}|/\epsilon_{\rm eff}\}$. The solid curves correspond to
Maxwell-Garnett theory as given by Eq.\ (\ref{plasmonlines}).}
\end{figure}

In a structure formed by a distribution of small nanoparticles,
one can use the Clausius-Mossotti formula \cite{J1975} that links
the polarizability of the particles $\alpha$ to the effective
dielectric function of such material $\epsilon_{\rm eff}$:
\begin{equation}\label{cm}
  \frac{\epsilon_{\rm
eff}-\epsilon_{\rm h}}{\epsilon_{\rm eff}+2 \epsilon_{\rm
h}}=\frac{4\pi}{3\epsilon_{\rm h}}\frac{\alpha}{v},
\end{equation}
where $v$ is the average volume per particle and $\epsilon_{\rm
h}$ is the dielectric function of the host material. This formula
does not take into consideration non-dipolar interactions, and the
effect of local order is neglected. However, it works quite well
for small particle concentrations. The polarizability of metallic
nanoparticles of radius $R$ can be obtained as a function of the
dielectric function of the metal $\epsilon_{\rm m}$ as
\begin{equation}\label{alpha}
\alpha=R^3 \epsilon_{\rm h} \frac{\epsilon_{\rm m}-\epsilon_{\rm
h}}{\epsilon_{\rm m}+2\epsilon_{\rm h}}.
\end{equation}
The combination of Eqs.\ (\ref{cm}) and (\ref{alpha}) is
equivalent to MGT. \cite{M1904} This theory predicts a single
plasmon for the composite medium that obeys the relation
\begin{equation}\label{plasmonlines}
\epsilon_{\rm m}=\frac{2(1-f)}{2+f} \epsilon_{\rm h}
\end{equation}
as a function of metal filling fraction $f=4\pi R^3/3 v$. Eq.\
(\ref{plasmonlines}) has been represented in Figs.\ \ref{Fig3} and
\ref{Fig4} by solid curves.

When the spheres are relatively close to each other, multipolar
terms become relevant in the effective response, which are not
accounted for by Eq.\ (\ref{plasmonlines}). Here, we have included
them by rigorously solving Maxwell's equations using the MESME
formalism sketched above and summarized in Eq.\ (\ref{eq9}), as
applied to periodic composites consisting of a large number of
layers of nanoparticles. First, reflectance and transmittance
matrices of incoming and scattered plane waves are obtained for each
layer using a MESME-like procedure. Then, the interaction among
layers is described by combining their reflectance and transmission
matrices. More details about this so-called layer-KKR method, along
with an efficient implementation, can be found in the excellent work
of Stefanou {\it et al.} \cite{SYM98} We have followed their
approach and have used less than 151 plane waves to obtain
convergence in the examples that follow, with the largest number of
waves needed when the spheres are nearly close-packet, as expected
from previously reported  divergence effects. \cite{PHW01}

The effective dielectric function $\epsilon_{\rm eff}$ is obtained
by comparison of the reflectance coefficient of a surface of the
composite materials under study to Fresnel's equations for
equivalent homogeneous media. \cite{J1975} We have found that a
single value of $\epsilon_{\rm eff}$ can reproduce the angular
dependence of the reflectance coefficients for all light
polarizations with excellent numerical precision as long as the
lattice period is small compared to the wavelength in the host
material.

\section{Results and discussion}

We have applied the above formalism to study the following
systems: periodic arrays of aluminum spheres placed in vacuum in
simple cubic and fcc configurations (Fig.\ \ref{Fig3}), and fcc
arrays of silver spheres surrounded by either vacuum or silicon
(Fig.\ \ref{Fig4}). The dielectric constants of the materials
under consideration have been taken from optical data.
\cite{P1985} In all cases, the distance between nearest neighbor
sites has been taken as 6 nm, although our results are quite
insensitive to the choice of this parameter, as long as it is much
smaller than the wavelength.

Figures\ \ref{Fig3} and \ref{Fig4} show contour plots of the
so-called loss function, ${\rm Im}\{-1/\epsilon_{\rm eff}\}$,
illustrating its dependence on metal filling fraction $f$ and
light frequency. In the absence of absorption, the loss function
diverges at the plasma frequencies ($\epsilon_{\rm eff}=0$), but
in the actual, lossy systems under discussion this divergence is
turned into finite peaks. Therefore, the regions of large values
of the loss function (bright areas) correspond to plasmon
excitations of the composites. Let us mention that the loss
function inherits its name from the relevant role that is plays in
electron energy loss spectroscopy (EELS), where it provides the
loss spectrum for electrons traversing the bulk of a material.
\cite{sam3}

Aluminum is a good example of a material where the Drude formula
(\ref{Drude}) describes extremely well its bulk dielectric
function with the parameters given in the caption of Fig.\
\ref{Fig2}. Therefore, we consider first a system of Al spheres in
vacuum as a prototype system to study plasmon chemistry in
metallic particle arrays. We observe that there are more than one
plasmon for each value of the filling fraction, unlike the single
plasmon prediction of Eq.\ (\ref{plasmonlines}) based upon MGT.
The plasmon frequency derived from that formula (solid curves in
Fig.\ \ref{Fig3}) follows relatively well the highest-frequency
plasmon of the rigorous calculation. This agreement is better in
the fcc lattice [Fig.\ \ref{Fig3}(b)]. However, this structure
exhibit other plasmons that originate in the splitting and mixing
of individual particle modes. These plasmon frequencies merge into
the isolated particle limit $\omega_p/\sqrt{3}$ as $f\rightarrow
0$. For small values of $f$, there are only two plasmon branches
that contribute significantly, arising from dipole-dipole
interaction among the spheres, although the low-frequency branch
is less clear in the simple cubic lattice. For larger $f$, new
plasmon modes show up in the loss function as a result of
interaction between multiples of higher order, in agreement with
Ref.\ \onlinecite{IPK04}. This trend becomes dramatic near the
percolation limit, where the material is expected to acquire good
metallic behavior. \cite{PM00,PEB01}

For an fcc array of Ag nanoparticles (Fig.\ \ref{Fig4}), one can
draw similar conclusions. When the host material is Si [Fig.\
\ref{Fig4}(b)] the plasmon peaks are red shifted as compared to
host vacuum [Fig.\ \ref{Fig4}(a)] and the plasmon peaks are
relatively more spaced in frequency for a given value of $f$.
Metallic nanoparticles embedded in Si constitute systems that have
become recently realizable in practice, \cite{FYB04} so that the
results presented here bear predictions that can be addressed
experimentally.

\begin{figure}
\includegraphics[keepaspectratio,width=6.0cm]{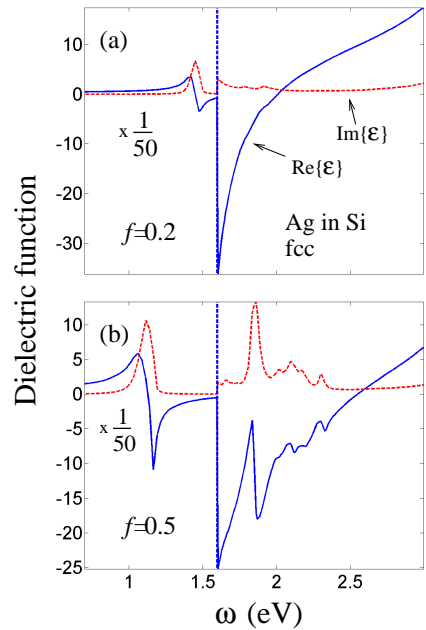}
\caption{\label{Fig5} Real and imaginary parts of the effective
dielectric function $\epsilon_{\rm eff}$ for an fcc arrangement of
Ag spheres surrounded by Si (solid and broken curves, respectively).
Two different filling fractions of the metal have been considered:
(a) $f=0.2$ and (b) $f=0.5$. The dielectric function has been
divided by 50 in the low-$\omega$ region to improve readability.}
\end{figure}

Both real and imaginary parts of the dielectric function have been
represented in Fig.\ \ref{Fig5} for fcc lattices of Ag spheres in
Si and with different values of the metal filling fraction
($f=0.2$ and $f=0.5)$. This figure indicates that the
highest-energy plasmon branch observed in Fig.\ \ref{Fig4}(b)
arises from to the vanishing of the real part of $\epsilon_{\rm
eff}$, with no further features showing up in the imaginary part,
which is consistent with the fact that this plasmon is relatively
close to the single plasmon predicted by MGT. This is quite
different from the lowest-energy plasmon, where $\epsilon_{\rm
eff}$ itself exhibits a resonant behavior translated into a
Lorentzian-type shape, including a peak in the imaginary part.
Other resonances of multipolar origin at intermediate energies
[see Fig.\ \ref{Fig4}(b)] are mainly associated to peaks in the
imaginary part, which are particularly clear in Fig.\
\ref{Fig5}(b) for $f=0.5$. Our calculations for the rest of the
combinations of lattice, metal, and embedding material considered
in this work allow us to draw a similar picture in those cases.

In conclusion, we have shown that the plasmons exhibited by
metallic nanoparticles do hybridize when the particles are close
to each other, and this produces new plasmon modes whose frequency
can be tuned by changing the distance between the particles. This
suggests the possibility of designing temperature and pressure
sensors based upon changes in the plasmon frequency with small
variations of the filling fraction, particularly near the
percolation limit (e.g., using short thiols to coat the particles
in order to have small separations between them). Our study
demonstrates the failure of Maxwell-Garnett theory,
\cite{M1904,LDL00} that predicts only a single plasmon in these
structures, rather than the rich structure of plasmons revealed by
our rigorous solution of the full electromagnetic problem, not to
mention Bruggeman's theory, \cite{B1935,LDL00} that does not
predict a plasmon at all. \cite{sam4}

\acknowledgments

FJGA would like to thank F. Meseguer, M. K\"all, J. M. Pitarke,
and J. Fern\'andez for helpful discussions. This work was
supported by the University of the Basque Country UPV/EHU
(contract No. 00206.215-13639/2001), by the late Spanish
Ministerio de Ciencia y Tecnolog\'{\i}a and the new Ministerio de
Educaci\'{o}n y Ciencia (contract Nos. MAT2001-0946 and
FIS2004-06490-C03-02, respectively), and by the European Union
({\it Metamorphose} Network of Excellence, NMP3-CT-2004-500252).
SR was partially supported by the Emil Aaltonen Foundation.


\end{document}